%% file: szoveg.tex
\documentclass[prd,onecolumn,nofootinbib,showpacs]{revtex4}
\usepackage{epsfig,graphicx,bm,feynmf}
\usepackage[latin2]{inputenc}
\usepackage{color}          
\usepackage{t1enc}
\usepackage{amssymb}
\usepackage{amsmath}

\newcommand{\tr}{\textnormal{Tr\,}}
\newcommand{\bea}{\begin{eqnarray}}
\newcommand{\eea}{\end{eqnarray}}  
\newcommand{\be}{\begin{equation}} 
\newcommand{\ee}{\end{equation}}

\newcommand{\p}{{\bf p}}

\newcommand{\no}{{\nonumber}}

\newcommand{\rt}[1]{{}}

\begin{document}

\title{
Resummed one--loop determination of the phase boundary of the
$SU(3)_R\times SU(3)_L$ linear sigma model in the $(m_\pi~-~m_K)$--plane
}

\author{T. Herpay}
\email{herpay@complex.elte.hu}
\affiliation{Department of Physics of Complex Systems,
E{\"o}tv{\"o}s University, H-1117 Budapest, Hungary}

\author{Zs. Sz{\'e}p}
\email{szepzs@achilles.elte.hu}
\affiliation{Research Institute for Solid State Physics and Optics
of the Hungarian Academy of Sciences, H-1525 Budapest, Hungary
}

\begin{abstract}
Complete one-loop parametrization of the linear sigma model is performed and 
the phase boundary between first order and crossover transition regions 
of the $m_\pi-m_K$--plane is determined using the optimized perturbation 
theory as a resummation tool of perturbative series. Away from the 
physical point the parameters of the model were determined by making 
use of chiral perturbation theory. Along the diagonal $m_\pi=m_K$ of 
the mass--plane we estimate $m_\pi^c=110\pm 20$ MeV.
The location of the tricritical point on the $m_\pi=0$
axis is estimated in the interval $m_K^{TCP}\in(1700,1850)$ MeV.
\end{abstract}

\maketitle

\section{Introduction}

In an attempt to understand the restoration of chiral and axial
$U(1)$ symmetries, chiral effective models are actively investigated
(see e.g. \cite{Roeder,Kalinovsky,Michalski} for some recent works). 
Effective models indicate a very rich
structure for the strongly interacting matter as function of quark
masses and various chemical potentials \cite{Barducci,Warringa}.
The effective treatment represents a 
complementary approach to the lattice QCD field theory which, 
however based on first principles, has difficulties mainly related to the 
computational power, in going towards the chiral limit $m_u=m_d=m_s=0$. 
These effective models are constructed to share the same global symmetries 
as the massless QCD. It is expected that the lower $m_u,m_d,m_s$ 
quark masses are (or alternatively $m_\pi$ and $m_K$) the better they work. 
Universal arguments \cite{pisarski84} predict a first order phase
transition for the chiral limit. Lattice simulations with staggered
quarks with a pion to rho mass ratio tuned to its physical value 
demonstrate a crossover type transition \cite{katz04}.
 
In QCD the critical line separating first order transitions from the 
crossover region in the $m_{u,d}-m_s$--plane is not precisely mapped,
because of the difficulties of simulating dynamical fermions.
There are several lattice studies with degenerate quarks
$m_u=m_d=m_s$, which show that the value of the pion mass on the
boundary between the crossover and first order phase transitions
drops substantially when finer lattices and improved actions are used,
from the initial estimates of $m_\pi^c\approx 290$ MeV
\cite{Karsch_high_mpc} or $m_\pi^c\approx 270$ MeV \cite{Christ} to
$m_\pi^c=67(18)$ MeV \cite{Karsch_low_mpc}. 
%or ever further down to $m_\pi^c< 65$ MeV \cite{MILC_low}.  
In view of such low values one hopes that the boundary of the phase 
transition can be investigated reliably using effective chiral models.

Although in principle it is simpler to solve an effective model than QCD, 
an exact solution cannot be given. Finding a good parametrization and
an adequate method of approximation are the key issues when dealing with them.
Attempts to parametrize physically the linear sigma model 
($L\sigma M$) date back to
the early 70's when in a series of papers Haymaker and
collaborators have performed it at tree-level
and started to calculate one-loop corrections at zero temperature
(see \cite{Haymaker73} and references therein). 
Recently other parametrizations were proposed in the literature 
\cite{Tornqvist99, Lenaghan00} (see also \cite{Herpay05}).  It turned out
that at tree-level it is not possible to fix the parametrization of the model
using only the well-known pseudoscalar masses, information is also 
needed from the less known scalar sector. Moreover, the
consequence of performing a tree-level parametrization 
is that one omits the effect of zero temperature vacuum fluctuations, 
which logarithmically depend on the renormalization scale. At finite 
temperature in the broken symmetry phase, the omitted terms have an 
additional implicit dependence on the temperature through the
masses which depends on the order parameter. 
If the effective model is solved in an approximation which is not
renormalization scale invariant, then the renormalization scale
appears as any other parameter of the theory, and it has to be included 
in the process of parametrization in which some quantities calculated
at quantum level are matched against their experimentally measured
physical values. The effect of the renormalization scale turns out to be 
both quantitatively  and qualitatively important. It can have an
effect on the pole structure of the scalar 
Green's function in the complex plane, as it happened in 
Ref.~\cite{Patkos04}. It influences the temperature dependence of the vacuum 
expectation value, and it can happen that above some temperature there 
is no solution to the equation of state (see e. g. \cite{Hatsuda98}).
The renormalization scale can even change the order of the phase transition.
All this reflects the approximate nature of the solution. A good idea is to 
choose a range of the renormalization scale where its variation 
affects the other parameters of the theory and the physical
quantities less (e.g. trying to achieve approximate renormalization scale 
independence).

Due to the effects of the renormalization scale mentioned above, it seems 
customary in the literature not to use the zero temperature quantum 
fluctuations of the field theory, and forgetting the renormalization 
issue just solve the model in a statistical mechanics inspired finite 
temperature quasi-particle approximation. 
Attempts to renormalize the model in the Hartree approximation of
the CJT-formalism \cite{CJT} were reported in 
\cite{Lenaghan00b}, but the result was not satisfactory.
Recently much effort has been put in the renormalization of
self-consistent resummation schemes of finite temperature QFT
\cite{Knoll,Reinosa,Borsanyi,Ivanov}. 
In view of these results, solving the model in a properly renormalized 
approximation is nowadays a compelling requirement. In the present
paper we want to go beyond the tree-level treatment of the model and its
quasi-particle thermodynamics as it was treated in \cite{Herpay05}, 
and investigate the challenges of solving the renormalized version of 
the model by taking into account the logarithmic corrections.  
In particular, we want to investigate the extent they influence the
location of the phase boundary in the pion--kaon mass--plane.

In section \ref{sec:param} we present the one-loop parametrization
of the model in the $m_\pi-m_K$--plane. It turns out to be  rather
hard to find a unique parametrization which works in the relevant region.
The thermodynamics and the influence of the logarithmic terms are
discussed at the physical point in section \ref{sec:thermo}. 
In section \ref{sec:boundary} we describe our results on the phase
boundary, and we conclude in section \ref{sec:conclusion}. 

\section{Parametrization of the model at one-loop level\label{sec:param}}

\noindent The Lagrangian of the $SU_L(3)~\times~SU_R(3)$ symmetric 
linear sigma model with  explicit symmetry breaking terms is given by
\begin{equation}
L(M)=\frac{1}{2}\tr(\partial_\mu M^{\dag} \partial^\mu M+\mu^2
M^{\dag} M)-f_1
\left( \tr(M^{\dag} M)\right)^2-f_2  \tr(M^{\dag}
M)^2-g\left(\det(M)+\det(M^{\dag})\right)+\epsilon_x\sigma_x+
\epsilon_y  \sigma_y,
\label{Lagrangian}
\end{equation}
where the mixing sector is written in the non-strange (x)- strange (y)
basis instead of the original 0 -- 8 basis, by performing an orthogonal 
transformation on the fields as in \cite{Herpay05} (see Appendix \ref{ap:A}). 
The complex 3$×$3 matrix $M$ defined by the scalar ($\sigma$)
and pseudoscalar ($\pi$) fields can be written as 
\begin{equation}
M=\frac{1}{\sqrt{2}}\sum_{i=1}^{7}(\sigma_i+i\pi_i)\lambda_i+\frac{1}{\sqrt{2}}
\textrm{diag}(\sigma_x+i\pi_x,\sigma_x+i\pi_x,\sqrt{2}(\sigma_y+i\pi_y)),
\label{xybasis}
\end{equation}
where $\lambda_i\,:\,\,\,i=1\ldots 7$ are the Gell-Mann matrices. 
Isospin breaking is not considered, therefore
in the broken phase only the scalar fields $\sigma_x$ and $\sigma_y$ 
have non-zero expectation values: $x:=\langle \sigma_x
\rangle$, $y:=\langle \sigma_y \rangle$.  
After shifting the fields in the Lagrangian by their expectation
values with a little bit of algebra one can perform the traces.
Details can be found in \cite{Haymaker73,Lenaghan00}.
Requiring that the sum of terms linear in the fluctuations vanishes
we obtain two equations of state. They are given explicitly in
section \ref{sec:thermo}. The coefficients of the quadratic
terms are the tree-level masses (see Tab.~\ref{Tab:masses}), while
the third and fourth order terms give the three- and four-point 
interaction vertices.

\begin{table}[!b]
%\begin{center}
\begin{tabular}{|l|l|}
\hline
$m^2_\pi\,\,\,\,\,=m^2+2(2f_1+f_2)x^2+4f_1y^2+2gy$ &
$m^2_{a_0}\,\,=m^2+2(2f_1+3f_2)x^2+4f_1y^2-2gy$ \\\hline
$m^2_K\,\,\,=m^2+2(2f_1+f_2)(x^2+y^2)+2f_2y^2-\sqrt{2}x(2f_2y-g)$
&$
 m^2_{\kappa}\,\,\,\,\,=m^2+2(2f_1+f_2)(x^2+y^2)+2f_2y^2+
\sqrt{2}x(2f_2y-g)$ 
\\\hline
$m^2_{\eta_{xx}}=m^2+2(2f_1+f_2)x^2+4f_1y^2-2gy$ & $
m^2_{\sigma_{xx}}=m^2+6(2f_1+f_2)x^2+4f_1y^2+2gy$ \\\hline
$ m^2_{\eta_{yy}}=m^2+4f_1x^2+4(f_1+f_2)y^2$ & $
m^2_{\sigma_{yy}}=m^2+4f_1x^2+12(f_1+f_2)y^2$ \\\hline
$m^2_{\eta_{xy}}=-2gx$ &$ m^2_{\sigma_{xy}}=8f_1xy+2gx $ \\\hline
\end{tabular}
\\
\caption{The squared masses of the (pseudo)scalar nonet appear in the
(first) second column. The last three rows represent the mixing sectors. 
They can be written in the conventional basis using the formulas of Appendix 
\ref{ap:A}.}
\vspace*{-0.25cm}
\label{Tab:masses}
\end{table}

In what follows, a set of non-linear one-loop equations will be given
which determines at $T=0$ the 8 parameters of the Lagrangian: the couplings 
$\mu$, $f_1$, $f_2$, $g$, the condensates $x$, $y$ and the
external fields $\epsilon_x$, $\epsilon_y$. 
Many ways of selecting these equations can be envisaged, see
\cite{Haymaker73} for alternatives. We have chosen to 
use as input the low lying pseudoscalar mass spectrum, namely the
pion, kaon and eta meson masses and the decay constants of the pion
and kaon, because they are the best known theoretically.

In the broken phase a resummation is need/home/szepzs/fermion/renorm/kor3/proof/DS9047-1.psed, in order to avoid the 
appearance of negative mass squares in the finite temperature 
calculations of one-loop quantities. This can be done for instance
using the Optimized Perturbation Theory (OPT) of Chiku and Hatsuda
\cite{Hatsuda98}. In the OPT the mass parameter $-\mu^2$ of the Lagrangian,
which in the broken phase could be negative, is replaced with an effective
(temperature-dependent) mass parameter $m^2$ which is determined 
using the criterion of fastest apparent convergence (FAC). 
The mass term of the Lagrangian reads:
\be
L_{mass}=\frac{1}{2}m^2 \tr M^\dag M-\frac{1}{2}(\mu^2+m^2) \tr M^\dag M
\equiv\frac{1}{2}m^2 \tr M^\dag M-\frac{1}{2} \Delta m^2\tr M^\dag M ,
\label{resum}
\ee
where the finite counterterm $\Delta m^2$
is taken into account first at one-loop level.

This resummation method replaces $-\mu^2$ by the effective mass
square $m^2$ in the tree-level masses (see Tab.~\ref{Tab:masses}), 
and preserves all the perturbative relations upon which Goldstone's 
theorem relies \cite{Hatsuda98}. The renormalization is achieved 
both in the symmetric and the broken phase by the following counterterms
\bea
\delta \mu^2 &= & \frac{   (5f_1+3f_2)\Lambda^2}{\pi^2}-  
\frac{(5f_1+3f_2)m^2-g^2}{\pi^2}\ln\frac{\Lambda^2}{l^2}, \label{mur} \\
\delta g&=&\frac{3g(f_1-f_2)}{2\pi^2}  \ln\frac{\Lambda^2}{l^2},  \\
\delta f_1&=&\frac{13f_1^2+12f_1 f_2+3f_2^2}{2\pi^2}
 \ln\frac{\Lambda^2}{l^2}, \\
\delta f_2&=&\frac{3f_1 f_2+3f_2^2}{\pi^2} \ln\frac{\Lambda^2}{l^2},  
\label{fkren}
\eea
where $\Lambda$ is the 3d regularization cutoff, and $l$ is the 
renormalization scale.
Note that only the mass counterterm differs from its standard expression 
\cite{Haymaker73}. In the present form this counterterm is
temperature-dependent through the effective mass, but this temperature 
dependence is canceled by higher-loop terms \cite{Hatsuda98,jako05}. 
In what follows, all quantities and equations are renormalized 
without any change in the notations.

The above mentioned FAC criterion, which determines the effective mass
square $m^2$, is realized in the present case by the requirement
that the pole and the residue of the one-loop pion propagator 
\be
D_\pi(p)=\frac{iZ_\pi^{-1}}{p^2-m_\pi^2-\Sigma_\pi(p^2,m_i,l)},
\label{pi_prop}
\ee
stay equal to their tree-level values. Here we anticipated that 
we also need a finite wave function renormalization in order to make 
the residuum equal to 1, and rescaled the pion fields as 
$\pi\to Z_\pi^{-\frac{1}{2}}\pi$.  

According to this FAC criterion the inverse of the finite wave function
renormalization constant is
\be
Z_\pi^{-1}:=1-\left.
\frac{\partial {\Sigma_\pi}(p^2,m_i,l) }{\partial p^2}
\right|_{p^2=M_\pi^2}.
\ee
The one-loop pion pole mass
\be
M_\pi^2=-\mu^2+(4{f_1}+2{f_2})x^2+4{f_1} y^2+2g y+
\textrm{Re}\left\{\Sigma_\pi(p^2=M_\pi^2,m_i,l)\right\} \label{Mpi}
\ee
has to be equal to its tree-level value 
($M_\pi\overset{\displaystyle!}{=}m_\pi$). 
Therefore, using the expression of the tree-level pion mass of
Tab.~\ref{Tab:masses}, the following ``gap'' equation can be obtained for the 
effective mass:
\be
m^2=-\mu^2+\textrm{Re}\left\{\Sigma_\pi(p^2=m_\pi^2,m_i(m^2),l)\right\},
\label{eqm}
\ee
where the $m^2$-dependence of the self-energy (through the tree-level masses) 
is explicitly shown. The different contributions to the self-energy
are depicted in Fig.~\ref{Fig:feyn}. 

The effective mass can be replaced by the pion mass by expressing it from
its tree-level formula. Then  
(\ref{eqm}) can be interpreted as a zero 
temperature gap-equation for the pion mass:
\be
m_\pi^2=-\mu^2+(4{f_1}+2{f_2})x^2+4{f_1} y^2+2g y+
\textrm{Re}\left\{\Sigma_\pi(p^2=m_\pi^2,m_i(m_\pi),l)\right\},
\label{pigap}
\ee
where the tree masses of all mesons
are expressed through the pion mass. A similar  
gap-equation will be used in the thermodynamical calculations for the 
temperature dependence of the pion mass. At $T=0$, the task is 
``reversed'': the pion mass is known and (\ref{pigap}) belongs to 
the set of equations, which determines the parameters. 
We have chosen to express the effective mass $m^2$ from the tree-level
mass formula of the pion because the pion has the smallest mass, and
positive solutions of (\ref{pigap}) ensure the positiveness of
all the other masses. We use the kaon and eta masses, the
relations of the Partially Conserved Axial-Vector Current (PCAC) for
the pion and kaon at one-loop order, and the two equations of state to
fix the remaining parameters.

\begin{figure}[!htpb]
\include{feyn}
\caption{The physical content of the one-loop pseudoscalar self-energies.}
\label{Fig:feyn}
\end{figure}

The one-loop kaon propagator is the following:
\be
D_K(p)=\frac{iZ_K^{-1}}{p^2-m_K^2-\Sigma_K(p^2,m_i,l)}.
\label{K_prop}
\ee
$Z_K^{-1}$ and the one-loop pole mass of the kaon $M_K$ can be calculated
similarly as in the case of the pion:
\be
Z_K^{-1}:=1-\left.
\frac{\partial {\Sigma_K}(p^2,m_i,l) }{\partial p^2}\right|_{p^2=M_K^2},
\ee
and
\be
M_K^2=-\mu^2+2(2f_1+f_2)(x^2+y^2)+2f_2y^2-\sqrt{2}x(2f_2y-g)+
\textrm{Re}\left\{\Sigma_K(p^2=M_K^2,m_i,l)\right\} .
\label{MK}
\ee

The description of the $\eta$ and $\eta^\prime$ mesons is slightly 
more complicated 
because of the mixing in the $x - y$ sector ( $0 - 8$ in the 
conventional basis).
The propagator is a 2$\times$2 matrix, and pole masses are defined
as the real part of the solutions of the following equations
\be
\textnormal{Det}\left.
\begin{pmatrix} 
p^2-m_{\eta_{xx}}^2-\Sigma_{\eta_{xx}}(p^2,m_i,l) \,\,& 
-m_{\eta_{xy}}^2-\Sigma_{\eta_{xy}}(p^2,m_i,l) \\ 
-m_{\eta_{xy}}^2-\Sigma_{\eta_{xy}}(p^2,m_i,l) & 
p^2-m_{\eta_{yy}}^2-\Sigma_{\eta_{yy}}(p^2,m_i,l) \,\,
\end{pmatrix} 
\right|_{p^2=M_\eta^2,M_{\eta^\prime}}=0 .
\ee
This yields two equations for the mass eigenvalues $M_\eta$, $M_{\eta^\prime}$:
\bea
M_{\eta}^2&=&\frac{1}{2}\textrm{Re}\left\{ m_{\eta_{xx}}^2+
\Sigma_{\eta_{xx}}(p^2=M_\eta^2,m_i,l)+m_{\eta_{yy}}^2+
\Sigma_{\eta_{yy}}(p^2=M_\eta^2,m_i,l) \right. \no \\
&-& \left.\sqrt{(m_{\eta_{xx}}^2+
\Sigma_{\eta_{xx}}(p^2=M_\eta^2,m_i,l)-m_{\eta_{yy}}^2-
\Sigma_{\eta_{yy}}(p^2=M_\eta^2,m_i,l))^2+4(m_{\eta_{xy}}^2+
\Sigma_{\eta_{xy}}(p^2=M_\eta^2,m_i,l) )^2 }\right\} ,
\label{Meta1}\\
M_{\eta^\prime}^2&=&\frac{1}{2}\textrm{Re}\left\{ m_{\eta_{xx}}^2+
\Sigma_{\eta_{xx}}(p^2=M_{\eta^\prime}^2,m_i,l)+m_{\eta_{yy}}^2+
\Sigma_{\eta_{yy}}(p^2=M_{\eta^\prime}^2,m_i,l) \right. \no \\
&+& \left.\sqrt{(m_{\eta_{xx}}^2+
\Sigma_{\eta_{xx}}(p^2=M_{\eta^\prime}^2,m_i,l)-m_{\eta_{yy}}^2-
\Sigma_{\eta_{yy}}(p^2=M_{\eta^\prime}^2,m_i,l))^2+4(m_{\eta_{xy}}^2+
\Sigma_{\eta_{xy}}(p^2=M_{\eta^\prime}^2,m_i,l) )^2 }\right\} .
\label{Meta2}
\eea

The definitions (\ref{Mpi}), (\ref{MK}), (\ref{Meta1}) and (\ref{Meta2})
of the pole masses give the correct one-loop masses only when
the self-energy is not complex. We note that if the tree-level masses
are close to their experimental values, then by looking at the
self-energy contributions of pion and kaon in Fig.~\ref{Fig:feyn}, 
one can recognize that they have no imaginary part for 
$p^2=m_\pi^2$ and $p^2=m_K^2$, respectively. This is also true in the
case of the $\eta$ self-energy. It turns out that the $\eta'$ self-energy 
has an imaginary part at the pole-mass determined as the zero of the real
part of the self-energy except the narrow range of $1820$ MeV $<l<1880$ MeV.
For this reason we decided not to include the one-loop equation 
for $M_{\eta^\prime}$ into the set of 
equations used for the parametrization.
We make up for the missing equation by extending FAC criterion
to the one-loop kaon mass too. This condition reads:
\be
M_K^2\overset{\displaystyle!}{=}m_K^2=m_\pi^2-2gy+4f_2y^2
-\sqrt{2}x(2f_2y-g). 
\label{mk}
\ee

Two more equations are provided by the one-loop PCAC relations
which according to \cite{Haymaker73} reads as
\bea
f_\pi&=&Z_\pi^{-\frac{1}{2}}\frac{-iD^{-1}_\pi(p=0)}{M_\pi^2}x, 
\label{PCAC_pi}
\\ 
f_K&=&Z_K^{-\frac{1}{2}}\frac{-iD^{-1}_K(p=0)}{M_K^2}
\frac{x+\sqrt{2}y}{2}.  
\label{PCAC_K}
\eea
As shown in Appendix \ref{ap:C}, these equations can be rewritten in an
explicitly renormalization scale-independent form.

Finally, the last two parameters, the two symmetry breaking external 
fields $\epsilon_x$ and $\epsilon_y$ are determined by the one-loop 
equations of state, with the help of zero temperature 
chiral Ward identities (Appendix \ref{ap:B}):
\bea
\epsilon_x &=& Z_\pi^{-1} \left(-iD^{-1}_\pi(p=0)\right) x, 
\label{xeq} \\
\epsilon_y &=& Z_K^{-1}  \left(-iD^{-1}_K(p=0)\right) 
\left(\frac{x}{\sqrt{2}}+y \right)-Z_\pi^{-1}\left(-iD^{-1}_\pi(p=0)\right) 
\frac{x}{\sqrt{2}} . 
\label{yeq} 
\eea

One can notice that, since OPT preserves Ward identities
at tree and at one-loop level as well, the above parametrization, 
in which the tree-level masses of pion and kaon equal the one-loop 
masses, ensures at zero temperature the validity of 
Goldstone's theorem for both pion and kaon.

\subsection{Parametrization at the physical point
\label{subsection:param}}

The parameters were determined as follows. From (\ref{Mpi}) and (\ref{mk}) 
one can express $\mu^2$ and $g$, respectively. Next, from the system of 4
non-linear equation (\ref{MK}), (\ref{Meta1}), (\ref{PCAC_pi}), 
(\ref{PCAC_K}) one can numerically determine $f_1$, $f_2$,
$x$, and $y$ as functions of the renormalization scale $l$. 
Going back to (\ref{Mpi}) and (\ref{mk}) one can compute
$\mu^2$ and $g$, respectively. Substituting
these parameters into (\ref{xeq}) and (\ref{yeq})
one can determine $\epsilon_x$ and $\epsilon_y$.
Unlike the tree-level parametrization case \cite{Herpay05},
now all scalar masses are predicted.

\begin{figure}[!t]
\includegraphics[keepaspectratio, width=0.6\textwidth]{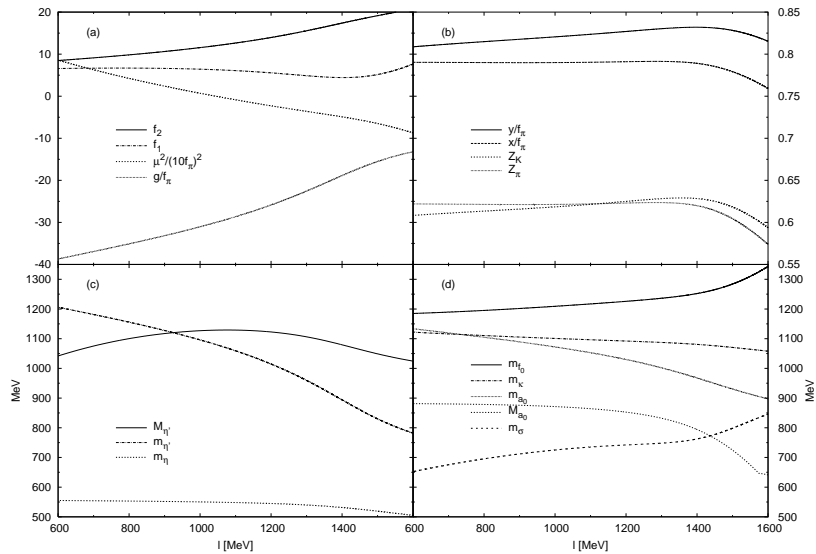}
\caption{
The renormalization scale dependence of various quantities at the
physical point: the parameters (a), the non-strange ($x$) and strange 
($y$) vacuum expectation values and the finite wave function
renormalization constants $Z_\pi$ and
$Z_K$ (b), the pseudoscalar masses (c), and the scalar masses (d).
%and of the estimation of one-loop pole-masses 
%based on the real parts of the self-energies.
}
\label{Fig:parametrisation}
\end{figure}

The numerical solution for different renormalization scales $l$ is presented 
in Fig.~\ref{Fig:parametrisation} (a) for the physical point, where 
$m_\pi=138$ MeV, $m_K=495.6$ MeV, $m_\eta=547.8$ MeV,
$f_\pi=93$ MeV, and $f_K=113$ MeV.
In figure Fig.~\ref{Fig:parametrisation} (b) one can see the
renormalization scale ($l$) dependence
of the non-strange ($x$) and strange ($y$) vacuum expectation values and
of the finite wave function renormalization constants $Z_\pi$ and $Z_K$.
They have a plateau for $l<1400$ MeV, and the tree-level 
$m_{\eta^\prime}$ (see Fig.~\ref{Fig:parametrisation} (c)) is the closest 
to its physical value in the region $l\in(1000,1400)$ MeV, where the 
variation of the tree-level scalar masses 
(see Fig.~\ref{Fig:parametrisation} (d)) is the mildest too. 
We have decided to use in our thermodynamical investigation
this range of the renormalization scale in which the tree-level masses
entering into the propagators of Fig.~\ref{Fig:feyn} are reasonably
close to their experimentally measured values.
In Fig.~\ref{Fig:parametrisation} (c, d)
we present an estimation of the predicted one-loop pole-masses based on the
real parts of the corresponding self-energies. This is a good approximation 
in the case of $\eta^\prime$, $a_0$ and $f_0$ since the zeros of
the inverse propagators correlate well with the location of the
well-defined peak in the corresponding spectral functions. The one-loop 
mass $M_{f_0}$, which is not shown in the figure, has a rather large value 
in the present range of the renormalization scale (decreasing 
from $M_{f_0}=2000$ MeV for $l=1000$ MeV to $M_{f_0}=1400$
MeV for $l=1400$ MeV). The shapes of the spectral functions of $\kappa$ 
and $\sigma$ (see Fig.~\ref{Fig:parametrisation} for $\rho_\sigma$) 
are more complicated, they have a threshold dominated peak with large width, 
and are very sensitive to the renormalization scale. In this case, it would 
be more appropriate to define the mass and width of a decaying particle
as  the real and imaginary part of a complex pole.
In the $O(N)$ model in the large N approximation \cite{Patkos02}, this
continuation into the second Riemann sheet was performed in the sigma
channel and the poles of the propagators were determined.
In this model the continuation of the propagators into the complex
plane would be more difficult due to the appearance of many decay 
thresholds and is beyond the scope of the present investigation. 

\subsection{Parametrization in the $m_\pi-m_K$--plane 
\label{subsection:elfolytatas}}

Since we are interested in the phase boundary on the $m_\pi-m_K$-plane, 
we have to take into account the variation of the 
parameters with $m_\pi$ and $m_K$. A method for the parametrization away from 
the physical point was proposed in \cite{Herpay05}, which relies on the 
formulas provided by the Chiral Perturbation Theory (ChPT) \cite{Gasser85}. 
Because our present parametrization does not use the $\eta^\prime$ meson, 
we make use of the $SU(3)$ ChPT describing the chiral dynamics of
the pseudoscalar octet. In the large $N_c$ limit, the formulas for
the pion, kaon mass dependence of the decay 
constants and of the $\eta$ mass up to $\mathcal{O}(1/f^2)$
read as \cite{Herrera98}: 
\bea
f_\pi&=&f\left(1+4L_5 \frac{m_\pi^2}{f^2}\right), \\
f_K&=&f\left(1+4L_5 \frac{m_K^2}{f^2}\right),  \\
m_\eta^2&=&\frac{4m_K^2-m_\pi^2}{3}+\frac{32}{3}(2L_8-L_5)
\frac{\left(m_K^2-m_\pi^2\right)^2}{f^2},
\label{Eq:eta_ChPT}
\eea 
where $L_5$ and $L_8$ are low energy constants and $f$ is the 
decay constant in the chiral limit. All the parameters of 
the large $N_c$ limit of the $SU(3)$ ChPT can be
determined at the physical point from the equations above.
Their values, $L_5=2.0152\cdot 10^{-3}$, $L_8=8.472\cdot 10^{-4}$ 
and $f=91.32$ MeV, are fixed for  all values of $m_\pi$ and $m_K$.

\begin{figure}[!t]
\centering{
\includegraphics[width=0.5\textwidth, keepaspectratio]{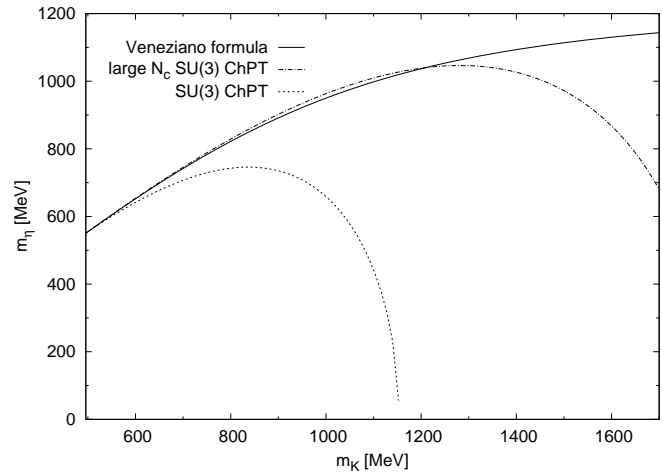}}
\caption{The kaon mass dependence of $m_\eta$ as given by ChPT and the 
Veneziano formula for $m_\pi=10$ MeV.}
\label{fig:Veneziano}
\end{figure}

For low values of $m_\pi$ and $m_K$ the sensitivity to the
renormalization scale of the $L\sigma M$ is bigger than the uncertainties 
coming from the omission of the chiral logarithms. As an effect of these
chiral logarithms, for large values of $m_K$, the formula
of the $SU(3)$ ChPT yields a decreasing value for $m_\eta$ for increasing
$m_K$, see Fig.~\ref{fig:Veneziano}.
If we use (\ref{Eq:eta_ChPT}), the same behavior occurs at a larger
value, which is around $m_K \approx 1300$ MeV. 
This is non-physical as both the kaon and the eta 
particles have to decouple in order to arrive at the O(4)
model for $m_K\to \infty$. This shows the failure of the ChPT at high 
values of the kaon mass.

In view of the bad behavior of $m_\eta$ determined using large $N_c$
ChPT, we have used, as an alternative, the following mass-formula by 
Veneziano
\cite{Veneziano} :
\be
m_\eta^2=m_K^2+\frac{1}{2}\Delta m_{\eta 0}^2-
\frac{1}{2}
\sqrt{\left(2m_K^2-2m_\pi^2-\frac{1}{3}\Delta m_{\eta 0}^2 \right)^2
+\frac{8}{9} \Delta m_{\eta 0}^4}.
\label{Eq:Veneziano}
\ee
$\Delta m_{\eta 0}^2$ is the non-perturbative gluonic mass contribution 
in the singlet channel of the mixing $\eta-\eta^\prime$ sector,
related to the axial $U(1)$ dynamics. 
Using the values of the masses at the physical point in (\ref{Eq:Veneziano})
one can fix the value of the extra mass contribution: 
$\Delta m_{\eta 0}^2=2.3$ GeV$^2$. One can see in Fig.~\ref{fig:Veneziano} that
this parametrization gives for $m_\eta$ 
values which are almost identical to the values coming from the
formula of ChPT in the large $N_c$ 
limit, up to values of $m_K$ for which ChPT breaks down. 

We note, that in the original paper  $\Delta m_{\eta 0}^2$ was
determined using the trace of the $2\times 2$ matrix of the mixing
$\eta-\eta^\prime$ sector. We indulged in modifying the procedure
in order to make contact with the ChPT in the large $N_c$ limit, as 
$m_\eta$ obtained form (\ref{Eq:Veneziano}) with the original
parametrization is always smaller than the value given by the large $N_c$ ChPT.

The continuation onto the $m_\pi-m_K$--plane of $f_\pi, f_K,$ and $m_\eta$,
based on the formulas of this subsection, allow us to determine the parameters
as described in \ref{subsection:param} in a wide region of the
mass-plane, except for high values of $m_K$, near the $m_\pi=0$ axis.

\section{Complete 1--loop thermodynamics of the L$\sigma$M\label{sec:thermo}}

With the intention of determining the order of the phase transition in 
the pion--kaon mass--plane we have to monitor the order parameters 
as functions of temperature. They are obtained from a set of three equations:
two equations of state for $x$ and $y$ and a gap-equation for the pion mass. 
The temperature dependence of the order parameters at finite T can be 
obtained from the equations of state:
\bea
-\epsilon_x+m^2 x+2g x y+4{f_{1}} x y^2+(4{f_{1}}+2{f_{2}})x^3+
\sum_{i} J_i t^x_i I(l,m_{i}(T),T) +\Delta m^2 x&=&0 \label{xeqT}\,,\\
-\epsilon_y+m^2 y+g x^2+4{f_{1}} x^2y+4({f_{1}}+{f_{2}})y^3+
\sum_{i} J_i t^y_i I(l,m_{i}(T),T)+\Delta m^2 y&=&0 \label{egy2T}\,, 
\eea
where in the tadpole integral $I(l,m_{i}(T),T)$ we have explicitly displayed 
the implicit temperature dependence of the 
tree-level masses, expressed with the pion mass determined by 
the gap-equation
\be
m_\pi^2=-\mu^2+(4{f_1}+2{f_2})x^2+4{f_1} y^2+2g y+
\Sigma_\pi(p^2=m_\pi^2,m_i(m_\pi),l)+\Sigma_\pi^T(\omega=0,m_i(m_\pi)). 
\label{pigapT}
\ee
The sum goes over all mass eigenstate meson fields
with isospin multiplicity factor $J_i$: $J_{\pi,a_0}=3$, 
$J_{K,\kappa}=4$, and  $J_{\eta,\eta',\sigma,f_0}=1$.
The coefficients
$t^x_i$ and $t^y_i$ appearing in (\ref{xeqT}) and 
(\ref{egy2T}) are listed in  Appendix C of \cite{Herpay05}. 
The standard one-loop integrals appearing in the formulas above can be
found for instance in \cite{Hatsuda98}. In the present study we use 
the one-loop bubble integrals appearing in (\ref{pigapT}) for
$|\p|=0$, corresponding to particles at rest. 
As in zero temperature case, at finite temperature OPT guarantees through the 
gap-equation the validity of Goldstone's theorem for the pion. 
Following Ref.~\cite{Hatsuda98}, $\Sigma_\pi^T$, the finite-temperature part 
of the self-energy, is taken not on the mass-shell, but instead at $\omega=0$.
This is done because above a given temperature 
$\Sigma_\pi(T)(\omega=m_\pi(T),\bf 0)$ becomes complex, and the real solution 
of the gap equation ceases to exist. At the physical mass-point this 
temperature is typically below the value of the pseudocritical
temperature, invalidating the study of the phase transition. 
The imaginary part is produced by one-loop bubble integrals
where two unequal masses $m_1$ and $m_2$ appear, when the relation 
$\omega=m_\pi<|m_1-m_2|$ is satisfied. It has the consequence that we
can not regard the pion gap-equation as the equation determining the 
``true'' one-loop mass of the pion. Due to the imaginary part of the 
self-energy the most adequate way of determining the pion mass would
be to look for a 
complex pole of the pion propagator. Still, in the present work we
content ourselves with the study of the spectral function
of the pion, which, as we will see, in certain temperature ranges
also provides information on the pion mass. 

\subsection{Influence of the logarithmic terms\label{subsec:influence}}

\begin{figure}[!h]
\centering{
\includegraphics[width=0.41\textwidth, keepaspectratio]{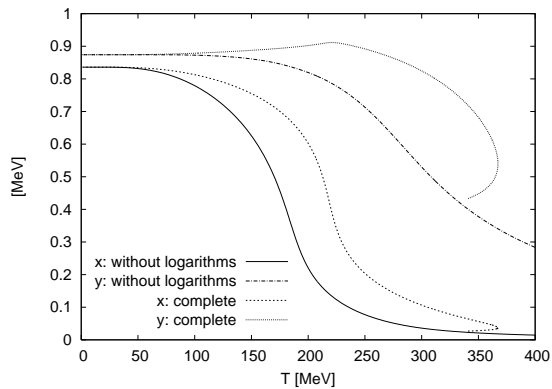}}
\vspace*{-0.2cm}
\caption{Comparison of the temperature dependence of strange ($y$) and
non-strange ($x$) condensates with and without the inclusion of the 
logarithms for $l=1200$ MeV.}
\label{fig:comparison}
\end{figure}

In order to estimate the influence of the logarithmic terms on the solution of 
the equations of state we have performed the following check. 
We have taken the 
parameters determined with the one-loop parametrization presented in 
Sec.~\ref{sec:param}, and modified the value of $\mu^2$ such as
to incorporate the zero temperature logarithmic terms. Then, at finite
temperature, we have solved the gap equation and the two equations of
state without taking into account the logarithmic terms, which implicitly 
depend on the temperature through the masses. The difference 
between these expectation values and the ones calculated 
with the logarithmic terms (and with the original,
unmodified value of $\mu^2$) is only due to the temperature dependence
of the logarithmic terms. This can be seen in Fig.~\ref{fig:comparison}: 
without the logarithmic terms the pseudocritical 
temperature is lower by about 20\%. The variation of strange
condensate with the temperature is significantly different in the two cases. 
We notice that at high temperature the solution obtained with
logarithmic term included ceases to exist, as was also observed in 
\cite{Hatsuda98}. This is a non-physical phenomenon, it happens
before the restoration of chiral symmetry completes. We consider the
solution reliable up to temperature values which are below the turning
point in $x$ and $y$.
\vspace*{-0.4cm}

\subsection{The solution at the physical point}

\begin{figure}[!tpb]
\centering{
\includegraphics[width=0.98\textwidth, keepaspectratio]{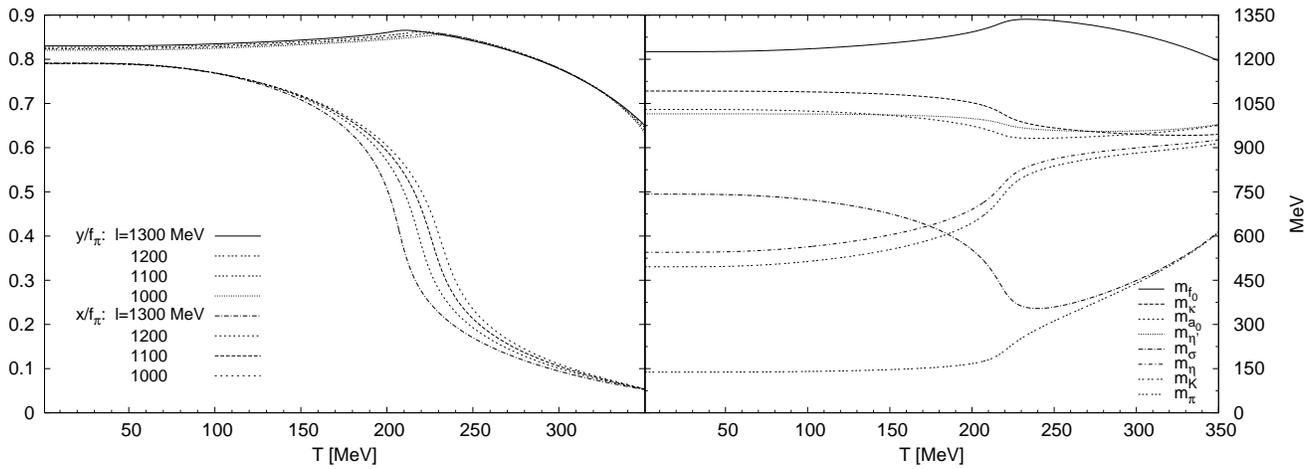}}
\caption{The temperature dependence of non-strange (x) and strange (y)
  order parameters for different values of the renormalization scale
(l.h.s) and the $T$-dependence of the tree-level masses
 for $l=1200$ MeV (r.h.s).}
\label{fig:mass_vev_T}
\end{figure}

At the physical point and in the investigated range of the renormalization 
scale, the behavior of the non-strange order parameter shows a smooth 
restoration of the $SU(2)$ chiral symmetry. The pseudocritical temperature 
moderately depends on the renormalization scale. The strange order 
parameter varies less and it is approximately scale-independent until 
the solution is reliable, see the left panel of Fig.~\ref{fig:mass_vev_T}. 
The right panel shows that the tree-level $SU(2)$ mass 
partners tend towards degeneracy as the temperature increases. 
Unfortunately the solution falls dead before the restoration of the complete 
$SU(3)$ symmetry, due to the effect of the logarithmic terms 
(see \ref{subsec:influence}). 

\begin{figure}[!b]
\centering{
\includegraphics[width=0.98\textwidth, keepaspectratio]{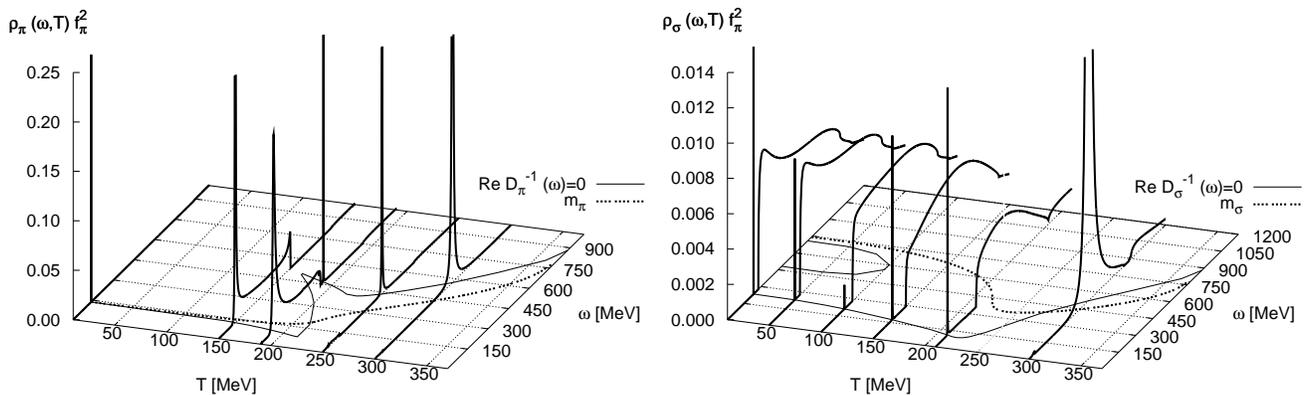}}
\caption{The spectral function of the pion (l.h.s.) and sigma (r.h.s.)
for various values of the temperature and $l=1200$ MeV.
The zeros of the real part of the inverse pion/sigma propagators 
and the corresponding tree-level masses are also depicted,
as lines in the $T-\omega$ plane.}
\label{fig:rho_pi_sigma_T}
\end{figure}

Fig.~\ref{fig:rho_pi_sigma_T} illustrates the temperature dependence
of the spectral functions in the pseudoscalar and scalar channels and the
behavior of the zeros of the real part of the inverse pion and sigma
propagators. At low temperature the spectral function in the pion
channel develops a peak whose location is close to the physical mass
value of the pion. Close to the pseudocritical temperature there are
significant changes in the peak structure. In this temperature range, 
based on the spectral function, one can not determine the true pion pole-mass. 
In this range, the tree-level pion mass, which is the solution of
the gap-equation (\ref{pigap}) interpolates between various
zeros of the inverse propagator, see the left panel of 
Fig.~\ref{fig:rho_pi_sigma_T}. At high temperature 
the spectral function has a well defined peak again, whose location 
correlates with the zero of the inverse pion propagator.

At low temperature the spectral function in the sigma channel has a
large width and the zeroes of the real part of the inverse propagator
are very sensitive to the renormalization scale, therefore the sigma pole 
mass can not be estimated. We can define the pole mass only at high 
temperature, where the spectral function develops a well--defined peak,
correlated with the zero of the inverse sigma propagator.
Increasing the temperature its location approaches the location of the 
peak in the pion channel and eventually they become degenerate. 
This shows that the one-loop masses of 
the pion and sigma also reflect the restoration of the 
$SU(2)$ chiral symmetry at high temperature.

\section{The phase boundary on the $m_\pi-m_K$--plane \label{sec:boundary}}

The main result of our work is the determination of the boundary
between the region where a crossover transition occurs, with a
smooth variation of the order parameters, and a first order
phase transition region, which is signaled by the multivaluedness of
the order parameters.

\begin{figure}[!b]
\centering{
\includegraphics[width=0.8\textwidth, keepaspectratio]{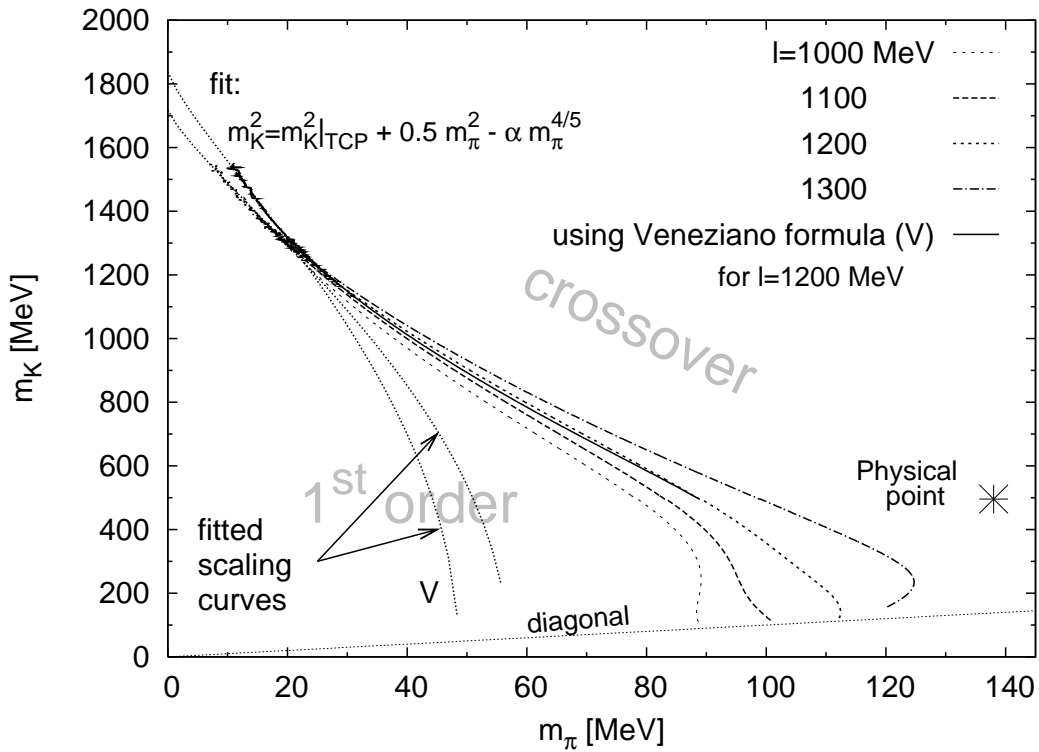}}
\caption{Phase boundary of L$\sigma$M obtained using a one-loop 
parametrization at $T=0$ and the formulas of large $N_c$ ChPT for
continuation on the $m_\pi-m_K$--plane. For large values of $m_K$ we
have also used the Veneziano formula for $m_\eta$ cf.
\ref{subsection:elfolytatas}.}
\label{fig:boundary}
\end{figure}

It is remarkable, that at large values of the kaon mass, the boundary
proves to be independent of the renormalization scale \footnote{
It can be seen
after some straightforward calculation, that as the kaon mass
increases, the relative weight of the terms containing
$\log(l)$ are becoming negligible in the equations
 used for parametrization as well as in those used for thermodynamical 
calculations.}. 
For large values of $m_K$ the existence of a scaling region belonging to
a tricritical point (TCP), with mean-field exponent can be confirmed. 
This is a new feature of the complete QFT treatment, it was not observed using 
tree-level parametrization! From mean-field studies (see
e.g. \cite{Hatta}) it is known that near the TCP the boundary of 
the edge of the first order region deviates from the $m_{u,d}=0$ axis
of the quark mass-plane according to $m_{u,d}\approx (m_s^{TCP}-m_s)^{5/2}$. 
Using the tree-level formulas of ChPT, for $m_\pi^2$ and $m_K^2$, namely
$m_\pi^2=2\hat m B_0$, $m_K^2=(\hat m+m_s)B_0$ with 
$\hat m=\frac{1}{2}(m_u+m_d)$, one can easily translate this into a
relation between the critical values of $m_K$ and $m_\pi$:
$m_K^2=m_K^2\big|_{TCP}+\frac{m_\pi^2}{2}-\alpha m_\pi^{4/5}$.
Due to the failure of our parametrization at very high values of
$m_K$, and near the $m_\pi=0$ axis we use this formula to extrapolate 
the upper edge of the phase boundary to the $m_K$-axis. Using large 
$N_c$ ChPT theory we obtain $m_K^{TCP}=1718$ MeV, while using 
the Veneziano formula (\ref{Eq:Veneziano}) we get
$m_K^{TCP}=1838$ MeV. Both fits are of very good quality suggesting
the radius of the scaling region to be $\Delta m_\pi^c\approx 40$ MeV. 
 In terms of the strange quark mass our estimate corresponds to
$m_s^{TCP}=13-15 \times m_s$.
This result can be compared to the recent lattice result of \cite{Philipsen}, 
where it was estimated that $m_s^{tric}\approx 3 m_s$.
In case of \cite{Philipsen} the piece of the phase boundary 
used in the extrapolation to $m_s^{TCP}$ was much closer to the physical
point and the location of the TCP was estimated using points 
with $m_s\le m_s^{phys}$. The lattice estimate for the location of
TCP would improve if points could be simulated closer to the scaling region.

In the present treatment the critical pseudoscalar mass for degenerate 
quarks was obtained in the range $m_\pi^c\in(90,130)$ MeV. The spread is
the result of the renormalization scale dependence. This range of $m_\pi^c$
is significantly above the one previously obtained in \cite{Herpay05} using a
tree-level parametrization, which eventually led to $m_\pi^c\in(20,60)$ MeV.
It is interesting to note, that the boundary lies to the right from
the fitted curves which describe the scaling behavior near TCP.

\section{Conclusion \label{sec:conclusion}}

In this paper we have studied the phase boundary of the linear sigma
model in the $m_\pi-m_K$ plane. We have managed to find in a resummed
perturbation theory a set of
renormalized equations based on which complete one-loop parametrization
of the model is possible not only at the physical point, but in a large
region of the mass-plane too. We allowed for the variation of the
parameters with $m_\pi$ and $m_K$ when moving away from the physical
point, and used formulas of the ChPT for the continuation of certain
physical quantities into the mass--plane.

As a result of the one-loop solution of the model we were able to
estimate the location of the tricritical point on the $m_\pi=0$ axis
of the mass plane at a rather high value of the kaon mass. We have shown
evidence of the existence of a scaling region around it.  These were not
observed previously in the quasi-particle approximations to the model.
For the same values of $m_K$ the phase boundary line lies at higher values of
pion masses than in the case of the tree-level parametrization of the
model, and without the inclusion of the vacuum fluctuations.  For
degenerate pseudoscalar masses the phase boundary was obtained in the
range $m_\pi^c\in(90,130)$ MeV.

We have discussed also the limitation of the optimized perturbation theory, 
related to the fact that one looks for real solutions of the gap-equation for
the resummed mass in temperature ranges where the self-energies develop
imaginary parts too. It would be interesting to study the
phase boundary using a self-consistent approximation based on the
exact propagator, like the 2PI approximation.

\section*{Acknowledgement}

Work supported by Hungarian Scientific Research Fund (OTKA) under
contract number T046129. Zs. Sz. is supported by OTKA Postdoctoral
Fellowship (grant no. PD 050015). We thank A. Patk{\'o}s for continuous
support and discussions during the work and the elaboration of the
paper.  We also thank P. Sz{\'e}pfalusy for useful comments.  We thank
P. Kov\'acs, A. Patk{\'o}s and P. Sz{\'e}pfalusy for carefully reading of the 
manuscript.

\appendix
\section{Connection between (0\,--\,8) and (x\,--\,y) basis \label{ap:A}}

From the $x-y$ basis used in section \ref{sec:param} one can obtain, 
with the help of an orthogonal transformation,
the mixed scalar-, pseudoscalar 
and the external fields in the conventional (0 -- 8) basis as: 
\be
\begin{pmatrix}\sigma_0\\ \sigma_8\end{pmatrix}:=O
\begin{pmatrix}\sigma_x\\ \sigma_y\end{pmatrix},
\begin{pmatrix}\pi_0\\ \pi_8\end{pmatrix}:=O\begin{pmatrix}\pi_x\\ \pi_y
\end{pmatrix}, \qquad
\begin{pmatrix}\epsilon_0\\ \epsilon_8\end{pmatrix}:=O
\begin{pmatrix}\epsilon_x\\ \epsilon_y\end{pmatrix}, \textnormal{ where  } O:=\frac{1}{\sqrt{3}} 
\begin{pmatrix}  \sqrt{2}   & \,\quad  1  \\  1\,     & -\sqrt{2}
\end{pmatrix} \, .
\label{Otrans}
\ee
The transformation of the mass matrices, and the mass eigenvalues are
\be
\begin{pmatrix}  m^2_{\eta_{00}} & m^2_{\eta_{08}} \\ 
m^2_{\eta_{08}} & m^2_{\eta_{88}} 
\end{pmatrix} =O 
\begin{pmatrix}  m^2_{\eta_{xx}} & m^2_{\eta_{xy}} \\ 
m^2_{\eta_{xy}} & m^2_{\eta_{yy}} 
\end{pmatrix} O\, ,\,\,\textnormal{ and} \quad 
m^2_{\eta^{'},\,\eta}=\frac{m^2_{\eta_{xx}}+m^2_{\eta_{yy}} \pm 
\sqrt{(m^2_{\eta_{xx}}-m^2_{\eta_{yy}})^2+4m^4_{\eta_{xy}}} }{2} \,.
\ee
Similar expressions also hold for the scalar fields.

\section{Ward identities\label{ap:B}}

From (\ref{pi_prop}) and (\ref{K_prop}), the inverse pion and kaon
propagators at zero external momenta are
\bea
-iZ_\pi^{-1} D^{-1}_\pi(p=0)&=&m_\pi^2+\Sigma_\pi(p=0)+\Delta m^2 \,
\\
-i Z_K^{-1} D^{-1}_K(p=0)&=&m_K^2+\Sigma_K(p=0)+\Delta m^2,
\eea
where the finite counterterms of OPT are now explicitly indicated. 
Comparing the expressions of tree-level masses
(Table~\ref{Tab:masses}) with the tree-level part of the equations of
state (\ref{xeqT}), (\ref{egy2T}), one can obtain the corresponding
tree-level Ward identities $\epsilon_x=m_\pi^2 x$ and
$\epsilon_y=m_K^2(\sqrt{2}/2x+y)-m_\pi^2\sqrt{2}/2x $. The diagrams
for the self-energies are shown in Fig.~\ref{Fig:feyn}. They include
both tadpole and bubble diagrams, and the latter ones can be decomposed
for non-equal propagator masses and zero external momenta into the 
difference of two tadpoles:
\begin{fmffile}{bubble}
\bea
\raisebox{-5.mm}{\begin{fmfchar*}(17,12)
        \fmfleft{i}
        \fmfright{o}
        \fmfforce{0.2w,0.5h}{v1}
        \fmfforce{0.8w,0.5h}{v2}
        \fmfforce{0.1w,0.85h}{v3}
	\fmfforce{-0.1w,0.5h}{eleje}
	\fmfforce{1.1w,0.5h}{vege}
       % \fmfv{d.sh=circle,d.fi=-.5,d.si=.15h}{v2}
        \fmf{dashes}{i,v1}
        \fmf{dashes}{o,v2}
        \fmf{dashes,left,label=$m_1$,label.dist=1mm,label.side=left}{v1,v2}
	\fmf{plain,right,label=$m_2$,label.dist=1.5mm,label.side=left}{v1,v2}
%	\fmf{phantom,label=$\pi$,label.dist=0.8mm,label.side=left}{eleje,v1}
%	\fmf{phantom,label=$\pi$,label.dist=0.8mm,label.side=left}{v2,vege}
       % \fmf{wiggly}{v1,v3}
      \end{fmfchar*}\ \ } \hspace*{-1mm}&=&\frac{1}{m_1^2-m_2^2} \cdot \Bigg[\,\,\, 
 \raisebox{-3mm}{\begin{fmfchar*}(13,13)
        \fmfforce{0.5w,0.7h}{top}
        \fmfforce{0.5w,0.0h}{v}
        \fmfforce{1.0w,0.0h}{right}
        \fmfforce{0.0w,0.0h}{left}
	\fmfforce{-0.1w,0.0h}{eleje}
	\fmfforce{1.1w,0.0h}{vege}	
	\fmfforce{0.0w,0.7h}{ie}
	\fmfforce{1.0w,0.7h}{iv}
        \fmf{dashes,left}{v,top,v}
        \fmf{dashes}{left,v}
	%\fmf{phantom,label=$\pi$,label.dist=0.8mm,label.side=left}{eleje,left}
%	\fmf{phantom,label=$\pi$,label.dist=0.8mm,label.side=left}{right,vege}
        \fmf{phantom,label=$m_1$,label.dist=-3mm,label.side=right}{ie,iv}
	\fmf{dashes}{right,v}
      \end{fmfchar*}\ \ } -\,\,\,
 \raisebox{-3mm}{\begin{fmfchar*}(13,13)
        \fmfforce{0.5w,0.7h}{top}
        \fmfforce{0.5w,0.0h}{v}
        \fmfforce{1.0w,0.0h}{right}
        \fmfforce{0.0w,0.0h}{left}
	\fmfforce{-0.1w,0.0h}{eleje}
	\fmfforce{1.1w,0.0h}{vege}	
	\fmfforce{0.0w,0.7h}{ie}
	\fmfforce{1.0w,0.7h}{iv}
        \fmf{plain,left}{v,top,v}
        \fmf{dashes}{left,v}
        \fmf{dashes}{right,v}
%	\fmf{phantom,label=$\pi$,label.dist=0.8mm,label.side=left}{eleje,left}
%	\fmf{phantom,label=$\pi$,label.dist=0.8mm,label.side=left}{right,vege}
        \fmf{phantom,label=$m_2$,label.dist=-3mm,label.side=right}{ie,iv}
      \end{fmfchar*}\ \ }
\Bigg]\,.
\eea
\end{fmffile}
Therefore, the pseudoscalar self energies at $p=0$ can be represented as
a linear combination of tadpole contributions whose weights are complicated
expressions of the tree-level masses and the corresponding four- and
three-point couplings (see \cite{Haymaker73}). In the case of pion and
kaon these weights simplify to
\bea
\Sigma_\pi(p=0)&=&\frac{\sum J_i t^x_{i} I(m_{i},l)+ x\Delta m^2}{x}\,,\\
\Sigma_K(p=0)&=&\frac{ \sum J_i (t^x_{i}+\sqrt{2}  t^y_{i})I(m_{i},l)+
(x +\sqrt{2}y)\Delta m^2   }{x +\sqrt{2}y}\,, 
\label{simplform}  
\eea
where $I(m_i,l)$ is the $T=0$ tadpole integral
 and the sum goes over all mass eigenstate meson fields
with isospin multiplicity $J_i$. Substituting these
and the corresponding tree-level Ward identities into the expressions
(\ref{pi_prop}), (\ref{K_prop}), one can obtain the equations of state
(\ref{xeqT}) and (\ref{egy2T}), which determine the external fields at
zero temperature in the parametrization process. The relations
(\ref{xeq}) and (\ref{yeq}) are also valid 
at finite temperature and ensure the fulfillment of Goldstone's
theorem at one-loop order.

\section{PCAC relations \label{ap:C}}

The one-loop order PCAC relations for the pion and kaon are given in
\cite{Haymaker73} by
\be
f_\pi M_\pi^2=\sqrt{Z_\pi}\epsilon_x \,, \quad\quad 
f_K M_K^2=\frac{\sqrt{Z_K}\epsilon_y}{\sqrt{2}} +\sqrt{Z_\pi}\epsilon_x \,.  
\ee
With the help of (\ref{xeq}), (\ref{yeq}) the external fields can be
eliminated and one can obtain the expressions (\ref{PCAC_pi}), 
(\ref{PCAC_K}), which appear renormalization 
scale-dependent. However, one can rearrange the pion self-energies 
appearing in (\ref{PCAC_pi}) as follows:
\bea
f_\pi=\frac{m_\pi^2+\Sigma_\pi(p=0,l)}{M_\pi^2}Z_\pi^{-\frac{1}{2}}&=&
\frac{m_\pi^2+\Sigma_\pi(p^2=M_\pi^2,l)+
(\Sigma_\pi(p=0,l)-\Sigma_\pi(p^2=M_\pi^2,l))}{M_\pi^2}Z_\pi^{-\frac{1}{2}} 
\no\\
&=&
\left(1-\frac{\tilde\Sigma_\pi(p^2=M_\pi^2)}{M_\pi^2}\right)
Z_\pi^{-\frac{1}{2}}. 
\label{inde}
\eea
In the last step above we used the fact, that
$m_\pi^2+\Sigma_\pi(p^2=M_\pi^2)$ is just the definition of the
pole mass $M_\pi^2$. 
The $p$-dependent part of the
self energy, $\tilde\Sigma$, like the wave function renormalization constant,
does not depend on the renormalization scale and in consequence 
(\ref{inde}) is actually scale-independent. 
The PCAC equation for the kaon can be analyzed in a similar way.

\end{document}

%% file: feyn.tex
\begin{fmffile}{grafok}
\bea
\Sigma_\pi&=&\hspace*{-0.4cm}\sum_{i=\pi,\,K,\,\eta,\,\eta^\prime} \hspace*{-0.3cm}
      \raisebox{0.5mm}{\begin{fmfchar*}(13,13)
        \fmfforce{0.5w,0.7h}{top}
        \fmfforce{0.5w,0.0h}{v}
        \fmfforce{1.0w,0.0h}{right}
        \fmfforce{0.0w,0.0h}{left}
	\fmfforce{-0.1w,0.0h}{eleje}
	\fmfforce{1.1w,0.0h}{vege}	
	\fmfforce{0.0w,0.7h}{ie}
	\fmfforce{1.0w,0.7h}{iv}
        \fmf{dashes,left}{v,top,v}
        \fmf{dashes}{left,v}
	\fmf{phantom,label=$\pi$,label.dist=0.8mm,label.side=left}{eleje,left}
	\fmf{phantom,label=$\pi$,label.dist=0.8mm,label.side=left}{right,vege}
        \fmf{phantom,label=$i$,label.dist=1mm,label.side=right}{ie,iv}
	\fmf{dashes}{right,v}
      \end{fmfchar*}\ \ }
+ \hspace*{-0.3cm} \sum_{i=a_0,\,\kappa,\,\sigma,\,f_0} \hspace*{-0.3cm} 
 \raisebox{0.5mm}{\begin{fmfchar*}(13,13)
        \fmfforce{0.5w,0.7h}{top}
        \fmfforce{0.5w,0.0h}{v}
        \fmfforce{1.0w,0.0h}{right}
        \fmfforce{0.0w,0.0h}{left}
	\fmfforce{-0.1w,0.0h}{eleje}
	\fmfforce{1.1w,0.0h}{vege}	
	\fmfforce{0.0w,0.7h}{ie}
	\fmfforce{1.0w,0.7h}{iv}
        \fmf{plain,left}{v,top,v}
        \fmf{dashes}{left,v}
        \fmf{dashes}{right,v}
	\fmf{phantom,label=$\pi$,label.dist=0.8mm,label.side=left}{eleje,left}
	\fmf{phantom,label=$\pi$,label.dist=0.8mm,label.side=left}{right,vege}
        \fmf{phantom,label=$i$,label.dist=1mm,label.side=right}{ie,iv}
      \end{fmfchar*}\ \ }
+\hspace*{-0.3cm} \sum_{i=a_0,\,\sigma,\,f_0} \hspace*{-0.3cm}\
     \raisebox{-5.5mm}{\begin{fmfchar*}(17,12)
        \fmfleft{i}
        \fmfright{o}
        \fmfforce{0.2w,0.5h}{v1}
        \fmfforce{0.8w,0.5h}{v2}
        \fmfforce{0.1w,0.85h}{v3}
	\fmfforce{-0.1w,0.5h}{eleje}
	\fmfforce{1.1w,0.5h}{vege}
       % \fmfv{d.sh=circle,d.fi=-.5,d.si=.15h}{v2}
        \fmf{dashes}{i,v1}
        \fmf{dashes}{o,v2}
        \fmf{dashes,left,label=$\pi$,label.dist=1mm,label.side=left}{v1,v2}
	\fmf{plain,right,label=$i$,label.dist=1mm,label.side=left}{v1,v2}
	\fmf{phantom,label=$\pi$,label.dist=0.8mm,label.side=left}{eleje,v1}
	\fmf{phantom,label=$\pi$,label.dist=0.8mm,label.side=left}{v2,vege}
       % \fmf{wiggly}{v1,v3}
      \end{fmfchar*}\ \ }
\hspace*{-0.03cm}+\hspace*{-0.01cm}\sum_{i=\eta,\,\eta^\prime}\hspace*{-0.1cm}
     \raisebox{-5.5mm}{\begin{fmfchar*}(17,12)
        \fmfleft{i}
        \fmfright{o}
        \fmfforce{0.2w,0.5h}{v1}
        \fmfforce{0.8w,0.5h}{v2}
        \fmfforce{0.1w,0.85h}{v3}
	\fmfforce{-0.1w,0.5h}{eleje}
	\fmfforce{1.1w,0.5h}{vege}
	
       % \fmfv{d.sh=circle,d.fi=-.5,d.si=.15h}{v2}
        \fmf{dashes}{i,v1}
        \fmf{dashes}{o,v2}
        \fmf{dashes,left,label=$i$,label.dist=1mm,label.side=left}{v1,v2}
	\fmf{plain,right,label=$a_0$,label.dist=1mm,label.side=left}{v1,v2}
	\fmf{phantom,label=$\pi$,label.dist=0.8mm,label.side=left}{eleje,v1}
	\fmf{phantom,label=$\pi$,label.dist=0.8mm,label.side=left}{v2,vege}
       % \fmf{wiggly}{v1,v3}
      \end{fmfchar*} } 
\hspace*{-0.00cm}+\hspace*{+0.1cm}
     \raisebox{-5.5mm}{\begin{fmfchar*}(17,12)
        \fmfleft{i}
        \fmfright{o}
        \fmfforce{0.2w,0.5h}{v1}
        \fmfforce{0.8w,0.5h}{v2}
        \fmfforce{0.1w,0.85h}{v3}
	\fmfforce{-0.1w,0.5h}{eleje}
	\fmfforce{1.1w,0.5h}{vege}
	
       % \fmfv{d.sh=circle,d.fi=-.5,d.si=.15h}{v2}
        \fmf{dashes}{i,v1}
        \fmf{dashes}{o,v2}
        \fmf{dashes,left,label=$K$,label.dist=1mm,label.side=left}{v1,v2}
	\fmf{plain,right,label=$\kappa$,label.dist=1mm,label.side=left}{v1,v2}
	\fmf{phantom,label=$\pi$,label.dist=0.8mm,label.side=left}{eleje,v1}
	\fmf{phantom,label=$\pi$,label.dist=0.8mm,label.side=left}{v2,vege}
       % \fmf{wiggly}{v1,v3}
      \end{fmfchar*} } 
     \hspace*{0.00cm}+\hspace*{+0.1cm}
     \raisebox{+0.23mm}{\begin{fmfgraph*}(17,1)
         \fmfforce{0.5w,0.0h}{v}
         \fmfforce{1.0w,0.0h}{right}
         \fmfforce{0.0w,0.0h}{left}
	 \fmfforce{-0.2w,0.0h}{eleje}
	 \fmfforce{1.2w,0.0h}{vege}	
	 \fmfforce{0.0w,0.0h}{ie}
	 \fmfforce{1.0w,0.0h}{iv}

         \fmf{dashes}{left,v}
	 \fmf{phantom,label=$\pi$,label.dist=0.8mm,label.side=left}{eleje,v}
	 \fmf{phantom,label=$\pi$,label.dist=0.8mm,label.side=left}{v,vege}
         \fmf{phantom,label=$\Delta m^2$,label.dist=2mm,label.side=right}{ie,iv}
	 \fmfv{d.sh=circle,d.fi=0.1,d.si=2.4h}{v}
	 \fmf{dashes}{right,v}
	 % \fmf{wiggly}{v1,v3}
      \end{fmfgraph*} } \no  \\
\Sigma_K&=&\hspace*{-0.4cm}\sum_{i=\pi,\,K,\,\eta,\,\eta^\prime} \hspace*{-0.3cm}
      \raisebox{0.5mm}{\begin{fmfchar*}(13,13)
        \fmfforce{0.5w,0.7h}{top}
        \fmfforce{0.5w,0.0h}{v}
        \fmfforce{1.0w,0.0h}{right}
        \fmfforce{0.0w,0.0h}{left}
	\fmfforce{-0.1w,0.0h}{eleje}
	\fmfforce{1.1w,0.0h}{vege}	
	\fmfforce{0.0w,0.7h}{ie}
	\fmfforce{1.0w,0.7h}{iv}
        \fmf{dashes,left}{v,top,v}
        \fmf{dashes}{left,v}
	\fmf{phantom,label=$K$,label.dist=0.8mm,label.side=left}{eleje,left}
	\fmf{phantom,label=$K$,label.dist=0.8mm,label.side=left}{right,vege}
        \fmf{phantom,label=$i$,label.dist=1mm,label.side=right}{ie,iv}
	\fmf{dashes}{right,v}
      \end{fmfchar*}\ \ }
+ \hspace*{-0.3cm} \sum_{i=a_0,\,\kappa,\,\sigma,\,f_0} \hspace*{-0.3cm} 
 \raisebox{0.5mm}{\begin{fmfchar*}(13,13)
        \fmfforce{0.5w,0.7h}{top}
        \fmfforce{0.5w,0.0h}{v}
        \fmfforce{1.0w,0.0h}{right}
        \fmfforce{0.0w,0.0h}{left}
	\fmfforce{-0.1w,0.0h}{eleje}
	\fmfforce{1.1w,0.0h}{vege}	
	\fmfforce{0.0w,0.7h}{ie}
	\fmfforce{1.0w,0.7h}{iv}
        \fmf{plain,left}{v,top,v}
        \fmf{dashes}{left,v}
        \fmf{dashes}{right,v}
	\fmf{phantom,label=$K$,label.dist=0.8mm,label.side=left}{eleje,left}
	\fmf{phantom,label=$K$,label.dist=0.8mm,label.side=left}{right,vege}
        \fmf{phantom,label=$i$,label.dist=1mm,label.side=right}{ie,iv}
      \end{fmfchar*}\ \ }
+\hspace*{-0.3cm} \sum_{i=a_0,\,\sigma,\,f_0} \hspace*{-0.3cm}\
     \raisebox{-5.5mm}{\begin{fmfchar*}(17,12)
        \fmfleft{i}
        \fmfright{o}
        \fmfforce{0.2w,0.5h}{v1}
        \fmfforce{0.8w,0.5h}{v2}
        \fmfforce{0.1w,0.85h}{v3}
	\fmfforce{-0.1w,0.5h}{eleje}
	\fmfforce{1.1w,0.5h}{vege}
       % \fmfv{d.sh=circle,d.fi=-.5,d.si=.15h}{v2}
        \fmf{dashes}{i,v1}
        \fmf{dashes}{o,v2}
        \fmf{dashes,left,label=$K$,label.dist=1mm,label.side=left}{v1,v2}
	\fmf{plain,right,label=$i$,label.dist=1mm,label.side=left}{v1,v2}
	\fmf{phantom,label=$K$,label.dist=0.8mm,label.side=left}{eleje,v1}
	\fmf{phantom,label=$K$,label.dist=0.8mm,label.side=left}{v2,vege}
       % \fmf{wiggly}{v1,v3}
      \end{fmfchar*}\ \ }
\hspace*{-0.03cm}+\hspace*{-0.15cm}\sum_{i=\pi,\,\eta,\,\eta^\prime}\hspace*{-0.1cm}
     \raisebox{-5.5mm}{\begin{fmfchar*}(17,12)
        \fmfleft{i}
        \fmfright{o}
        \fmfforce{0.2w,0.5h}{v1}
        \fmfforce{0.8w,0.5h}{v2}
        \fmfforce{0.1w,0.85h}{v3}
	\fmfforce{-0.1w,0.5h}{eleje}
	\fmfforce{1.1w,0.5h}{vege}
	
     % \fmfv{d.sh=circle,d.fi=-.5,d.si=.15h}{v2}
        \fmf{dashes}{i,v1}
        \fmf{dashes}{o,v2}
        \fmf{dashes,left,label=$i$,label.dist=1mm,label.side=left}{v1,v2}
	\fmf{plain,right,label=$\kappa$,label.dist=1mm,label.side=left}{v1,v2}
	\fmf{phantom,label=$K$,label.dist=0.8mm,label.side=left}{eleje,v1}
	\fmf{phantom,label=$K$,label.dist=0.8mm,label.side=left}{v2,vege}
       % \fmf{wiggly}{v1,v3}
      \end{fmfchar*} }
     \hspace*{0.00cm}+\hspace*{+0.1cm}
     \raisebox{+0.23mm}{\begin{fmfgraph*}(17,1)
         \fmfforce{0.5w,0.0h}{v}
         \fmfforce{1.0w,0.0h}{right}
         \fmfforce{0.0w,0.0h}{left}
	 \fmfforce{-0.2w,0.0h}{eleje}
	 \fmfforce{1.2w,0.0h}{vege}	
	 \fmfforce{0.0w,0.0h}{ie}
	 \fmfforce{1.0w,0.0h}{iv}

         \fmf{dashes}{left,v}
	 \fmf{phantom,label=$K$,label.dist=0.8mm,label.side=left}{eleje,v}
	 \fmf{phantom,label=$K$,label.dist=0.8mm,label.side=left}{v,vege}
         \fmf{phantom,label=$\Delta m^2$,label.dist=2mm,label.side=right}{ie,iv}
	 \fmfv{d.sh=circle,d.fi=0.1,d.si=2.4h}{v}
	 \fmf{dashes}{right,v}
	 % \fmf{wiggly}{v1,v3}
      \end{fmfgraph*} } \no \\
\Sigma_{\,\eta_{kl}}\hspace*{-0.2cm}&=&\hspace*{-0.4cm}\sum_{i=K,\,\eta,\,\eta^\prime} \hspace*{-0.1cm}
      \raisebox{0.5mm}{\begin{fmfchar*}(13,13)
        \fmfforce{0.5w,0.7h}{top}
        \fmfforce{0.5w,0.0h}{v}
        \fmfforce{1.0w,0.0h}{right}
        \fmfforce{0.0w,0.0h}{left}
	\fmfforce{-0.1w,0.0h}{eleje}
	\fmfforce{1.1w,0.0h}{vege}	
	\fmfforce{0.0w,0.7h}{ie}
	\fmfforce{1.0w,0.7h}{iv}
        \fmf{dashes,left}{v,top,v}
        \fmf{dashes}{left,v}
	\fmf{phantom,label=$k$,label.dist=0.8mm,label.side=left}{eleje,left}
	\fmf{phantom,label=$l$,label.dist=0.8mm,label.side=left}{right,vege}
        \fmf{phantom,label=$i$,label.dist=1mm,label.side=right}{ie,iv}
	\fmf{dashes}{right,v}
      \end{fmfchar*}\ \ }
+ \hspace*{2mm}
 \raisebox{0.5mm}{\begin{fmfchar*}(13,13)
        \fmfforce{0.5w,0.7h}{top}
        \fmfforce{0.5w,0.0h}{v}
        \fmfforce{1.0w,0.0h}{right}
        \fmfforce{0.0w,0.0h}{left}
	\fmfforce{-0.1w,0.0h}{eleje}
	\fmfforce{1.1w,0.0h}{vege}	
	\fmfforce{0.0w,0.7h}{ie}
	\fmfforce{1.0w,0.7h}{iv}
        \fmf{plain,left}{v,top,v}
        \fmf{dashes}{left,v}
        \fmf{dashes}{right,v}
	\fmf{phantom,label=$k$,label.dist=0.8mm,label.side=left}{eleje,left}
	\fmf{phantom,label=$l$,label.dist=0.8mm,label.side=left}{right,vege}
        \fmf{phantom,label=$\kappa$,label.dist=1mm,label.side=right}{ie,iv}
      \end{fmfchar*}\ \ }
+\hspace*{-0.1cm} \sum_{i=\sigma,\,f_0}^{j=\eta,\,\eta^\prime} \hspace*{-0.18cm}\
     \raisebox{-5.5mm}{\begin{fmfchar*}(17,12)
        \fmfleft{i}
        \fmfright{o}
        \fmfforce{0.2w,0.5h}{v1}
        \fmfforce{0.8w,0.5h}{v2}
        \fmfforce{0.1w,0.85h}{v3}
	\fmfforce{-0.1w,0.5h}{eleje}
	\fmfforce{1.1w,0.5h}{vege}
       % \fmfv{d.sh=circle,d.fi=-.5,d.si=.15h}{v2}
        \fmf{dashes}{i,v1}
        \fmf{dashes}{o,v2}
        \fmf{dashes,left,label=$j$,label.dist=1mm,label.side=left}{v1,v2}
	\fmf{plain,right,label=$i$,label.dist=1mm,label.side=left}{v1,v2}
	\fmf{phantom,label=$k$,label.dist=0.8mm,label.side=left}{eleje,v1}
	\fmf{phantom,label=$l$,label.dist=0.8mm,label.side=left}{v2,vege}
       % \fmf{wiggly}{v1,v3}
      \end{fmfchar*}\ \ }
\hspace*{-0.1cm}+\hspace*{0.1cm}
     \raisebox{-5.5mm}{\begin{fmfchar*}(17,12)
        \fmfleft{i}
        \fmfright{o}
        \fmfforce{0.2w,0.5h}{v1}
        \fmfforce{0.8w,0.5h}{v2}
        \fmfforce{0.1w,0.85h}{v3}
	\fmfforce{-0.1w,0.5h}{eleje}
	\fmfforce{1.1w,0.5h}{vege}
	
       % \fmfv{d.sh=circle,d.fi=-.5,d.si=.15h}{v2}
        \fmf{dashes}{i,v1}
        \fmf{dashes}{o,v2}
        \fmf{dashes,left,label=$\pi$,label.dist=1mm,label.side=left}{v1,v2}
	\fmf{plain,right,label=$a_0$,label.dist=1mm,label.side=left}{v1,v2}
	\fmf{phantom,label=$k$,label.dist=0.8mm,label.side=left}{eleje,v1}
	\fmf{phantom,label=$l$,label.dist=0.8mm,label.side=left}{v2,vege}
       % \fmf{wiggly}{v1,v3}
      \end{fmfchar*} } 
\hspace*{-0.00cm}+\hspace*{+0.1cm}
     \raisebox{-5.5mm}{\begin{fmfchar*}(17,12)
        \fmfleft{i}
        \fmfright{o}
        \fmfforce{0.2w,0.5h}{v1}
        \fmfforce{0.8w,0.5h}{v2}
        \fmfforce{0.1w,0.85h}{v3}
	\fmfforce{-0.1w,0.5h}{eleje}
	\fmfforce{1.1w,0.5h}{vege}
	
       % \fmfv{d.sh=circle,d.fi=-.5,d.si=.15h}{v2}
        \fmf{dashes}{i,v1}
        \fmf{dashes}{o,v2}
        \fmf{dashes,left,label=$K$,label.dist=1mm,label.side=left}{v1,v2}
	\fmf{plain,right,label=$\kappa$,label.dist=1mm,label.side=left}{v1,v2}
	\fmf{phantom,label=$k$,label.dist=0.8mm,label.side=left}{eleje,v1}
	\fmf{phantom,label=$l$,label.dist=0.8mm,label.side=left}{v2,vege}
       % \fmf{wiggly}{v1,v3}
      \end{fmfchar*} } \no \\
    && +\, \delta_{kl} \Bigg[\,\,\,
       \raisebox{-0.mm}{\begin{fmfchar*}(13,13)
        \fmfforce{0.5w,0.7h}{top}
        \fmfforce{0.5w,0.0h}{v}
        \fmfforce{1.0w,0.0h}{right}
        \fmfforce{0.0w,0.0h}{left}
	\fmfforce{-0.1w,0.0h}{eleje}
	\fmfforce{1.1w,0.0h}{vege}	
	\fmfforce{0.0w,0.7h}{ie}
	\fmfforce{1.0w,0.7h}{iv}
        \fmf{dashes,left}{v,top,v}
        \fmf{dashes}{left,v}
	\fmf{phantom,label=$k$,label.dist=0.8mm,label.side=left}{eleje,left}
	\fmf{phantom,label=$l$,label.dist=0.8mm,label.side=left}{right,vege}
        \fmf{phantom,label=$\pi$,label.dist=1mm,label.side=right}{ie,iv}
	\fmf{dashes}{right,v}
      \end{fmfchar*}  } 
       \hspace*{0.1cm}+ \hspace*{-0.4cm} \sum_{i=a_0,\,\sigma,\,f_0} \hspace*{-0.2cm} 
 \raisebox{0.5mm}{\begin{fmfchar*}(13,13)
        \fmfforce{0.5w,0.7h}{top}
        \fmfforce{0.5w,0.0h}{v}
        \fmfforce{1.0w,0.0h}{right}
        \fmfforce{0.0w,0.0h}{left}
	\fmfforce{-0.1w,0.0h}{eleje}
	\fmfforce{1.1w,0.0h}{vege}	
	\fmfforce{0.0w,0.7h}{ie}
	\fmfforce{1.0w,0.7h}{iv}
        \fmf{plain,left}{v,top,v}
        \fmf{dashes}{left,v}
        \fmf{dashes}{right,v}
	\fmf{phantom,label=$k$,label.dist=0.8mm,label.side=left}{eleje,left}
	\fmf{phantom,label=$l$,label.dist=0.8mm,label.side=left}{right,vege}
        \fmf{phantom,label=$i$,label.dist=1mm,label.side=right}{ie,iv}
      \end{fmfchar*}\ \ }
  \hspace*{0.00cm}+\hspace*{+0.1cm}
     \raisebox{+0.23mm}{\begin{fmfgraph*}(17,1)
         \fmfforce{0.5w,0.0h}{v}
         \fmfforce{1.0w,0.0h}{right}
         \fmfforce{0.0w,0.0h}{left}
	 \fmfforce{-0.2w,0.0h}{eleje}
	 \fmfforce{1.2w,0.0h}{vege}	
	 \fmfforce{0.0w,0.0h}{ie}
	 \fmfforce{1.0w,0.0h}{iv}

         \fmf{dashes}{left,v}
	 \fmf{phantom,label=$k$,label.dist=0.8mm,label.side=left}{eleje,v}
	 \fmf{phantom,label=$l$,label.dist=0.8mm,label.side=left}{v,vege}
         \fmf{phantom,label=$\Delta m^2$,label.dist=2mm,label.side=right}{ie,iv}
	 \fmfv{d.sh=circle,d.fi=0.1,d.si=2.4h}{v}
	 \fmf{dashes}{right,v}
	 % \fmf{wiggly}{v1,v3}
     \end{fmfgraph*} }  \,\Bigg] 
%\quad \quad \textnormal{ \parbox[l]{8cm}{{\bf{Figure I.}} This figure shows, what kind of diagram appears in the pseudoscalar self-energies} } 
\no 
\eea
\end{fmffile}

%% file: szoveg.bbl
\begin{thebibliography}{99}
\bibitem{Roeder}
  D. R{\"o}der, J. Ruppert and D. H. Rischke,
  Phys. Rev. D {\bf 68}, 016003 (2003)
  [nucl-th/0301085].
\bibitem{Kalinovsky}
  P. Costa, M. C. Ruivo, C. A de Sousa and Yu. L. Kalinovsky,
  Phys. Rev.  D {\bf 71}, 116002 (2005)
  [hep-ph/0503258].
\bibitem{Michalski}
  S. Michalski, [hep-ph/0601255].
\bibitem{Barducci}
  A. Barducci, R. Casalbuoni, G. Pettini, L. Ravagli,
  Phys. Rev. D {\bf 69}, 096004 (2004)
  [hep-ph/0402104].
\bibitem{Warringa}
  H. J. Warringa, D. Boer, J. O. Andersen
  Phys. Rev. D {\bf 72}, 014015 (2005)
  [hep-ph/0504177].
\bibitem{pisarski84}
  R. D. Pisarski and F. Wilczek,
  Phys. Rev. D {\bf 29}, 334 (1984).
\bibitem{katz04}
  Z. Fodor, S.D. Katz,
  JHEP 0404 (2004) 050
  [hep-lat/0402006].
\bibitem{Karsch_high_mpc}
  F. Karsch, E. Laermann, Ch. Schmidt,
  Phys. Lett. {\bf B520}, 41 (2001)
  [hep-lat/0107020].
\bibitem{Christ}
  N. H. Christ, X. Liao,
  Nucl. Phys. Proc. Suppl. {\bf 119}, 514 (2003).
\bibitem{Karsch_low_mpc}
  F. Karsch, C. R. Allton, S. Ejiri, S. J. Hands, O. Kaczmarek,
  E. Laermann and C. Schmidt,
  Nucl. Phys. Proc. Suppl. {\bf 129-130}, 614 (2004)
  [hep-lat/0309116].
%\bibitem{MILC_low}
%  The MILC Collaboration: C. Bernard, T. Burch, C. DeTar, S. Gottlieb,
%  E.B. Gregory, U.M. Heller, J. Osborn, R.L. Sugar, D. Toussaint
%  [hep-lat/0409097].
\bibitem{Haymaker73}
  L.\--H.\ Chan and R.\ W.\ Haymaker,
  Phys.\ Rev.\ D {\bf 7}, 402 (1973)
  [nucl-th/9901049].
\bibitem{Tornqvist99}
  N. A. T{\"o}rnqvist,
  Eur. Phys. J C {\bf 11}, 359 (1999)
  [hep-ph/9905282].
\bibitem{Lenaghan00}
  J.T. Lenaghan, D. H. Rischke and J. Schaffner-Bielich,
  Phys. Rev. D {\bf 62}, 085008 (2000)
  [nucl-th/0004006]. 
\bibitem{Herpay05}
  T. Herpay, A. Patk{\'o}s, Zs. Sz{\'e}p, P. Sz{\'e}pfalusy
  Phys. Rev. D {\bf 71}, 125017 (2005)
  [hep-ph/0504167].
\bibitem{Patkos04}
  A. Jakov\'ac, A. Patk{\'o}s,  Zs. Sz{\'e}p, P. Sz{\'e}pfalusy
  Phys. Lett. B {\bf 582}, 179 (2004)
  [hep-ph/0312088].  
\bibitem{Hatsuda98}
  S. Chiku and T. Hatsuda,
  Phys. Rev. D {\bf 58}, 076001 (1998)
  [hep-ph/9803226].
\bibitem{CJT}
  J. M. Cornwall, R. Jackiw, and E. Tomboulis,
  Phys. Rev. D {\bf 10}, 2428 (1974).
\bibitem{Lenaghan00b}
  J. T. Lenaghan, D. H. Rischke, 
  J. Phys. {\bf G26}, 431 (2000)
  [nucl-th/9901049].
\bibitem{Knoll}
  H. Hees, J. Knoll,
  Phys. Rev. D65, 025010 (2002)
  [hep-ph/0107200].
\bibitem{Reinosa}
  J.-P. Blaizot, E. Iancu, U. Reinosa
  Phys. Lett. {\bf B 568}, 160 (2003) 
  [hep-ph/0301201]; 
  Nucl. Phys. {\bf A 736}, 149  (2004)
  [hep-ph/0312085].
\bibitem{Borsanyi}
  J. Berges, Sz. Bors\'anyi, U. Reinosa, J. Serreau,
  Annals Phys. {\bf 320}, 344 (2005)
  [hep-ph/0503240]. 
\bibitem{Ivanov}
  Yu. B. Ivanov, F. Riek, H. van Hees, J. Knoll 
  Phys. Rev. D {\bf 72} (2005) 036008
  [hep-ph/0506157].
\bibitem{jako05}                              
  A. Jakov\'ac and Zs. Sz{\'e}p, 
  Phys.\ Rev.\ D {\bf 71}, 105001 (2005)
  [hep-ph/0405226].
\bibitem{Patkos02}
  A. Patk{\'o}s, Zs. Sz{\'e}p, P. Sz{\'e}pfalusy
  Phys. Rev. D 66 (2002) 116004
  [hep-ph/0206040].
\bibitem{Gasser85}
  J. Gasser, H. Leutwyler,
  Nucl. Phys. {\bf B250}, 465 (1985).
\bibitem{Herrera98}
  P. Herrera-Sikl{\'o}dy, J. I. Latorre, P. Pascual, J. Taron,
  Phys. Lett. B {\bf 419}, 326 (1998)
  [hep-ph/9710268].
\bibitem{Veneziano}
  G. Veneziano,
  Nucl. Phys. {\bf B159}, 213 (1979).
\bibitem{Hatta}
  Y. Hatta, T. Ikeda,
  Phys. Rev. D {\bf 67}, 014028 (2003)
  [hep-ph/0210284].
\bibitem{Philipsen}
  O. Philipsen,
  PoS LAT2005 (2005) 016
  [hep-lat/0510077].

\end{thebibliography}
